# Can Voters Detect Errors on Their Printed Ballots? Absolutely.

Philip Kortum, Michael D. Byrne, Chidera O. Azubike, Laura E. Roty
Department of Psychological Sciences
Rice University

**Abstract**

There is still debate on whether voters can detect malicious changes in their printed ballot after making their selections on a Ballot Marking Device (BMD). In this study, we altered votes on a voter's ballot after they had made their selections on a BMD. We then required them to examine their ballots for any changes from the slate they used to vote. Overall accuracy was exceptionally high. Participants saw 1440 total contests, and of those 1440, there were a total of 4 errors, so total accuracy was 99.8%. Participants were able to perform with near-perfect accuracy regardless of ballot length, ballot type, number of altered races, and location of altered races. Detection performance was extremely robust. We conclude that with proper direction and resources, voters can be near-perfect detectors of ballot changes on printed paper ballots after voting with a BMD. This finding has significant implications for the voting community as BMD use continues to grow. Research should now focus on identifying administrative and behavioral methods that will prompt and encourage voters to check their BMD-generated ballots before they drop them in the ballot box.

**Introduction**

The last several years have been difficult ones for American democracy. A large number of Americans continue to question the overall integrity of our voting systems, and their confidence that their votes will be accurately counted is the lowest seen in the last 15 years (McCarthy, 2020). The 2020 presidential election brought accusations that voting systems had been compromised and election results altered by malicious voting machines. To date, these claims have been rejected by the Department of Justice, the Election Assistance Commission, the courts, Election officials, and election experts (Brennan Center for Justice, 2020), but voting confidence remains shaken, with only 56% of Americans currently expressing confidence in the ability of our voting systems to reflect the will of the voters (Agiesta, 2022).

This is troubling, particularly since the United States has made a concerted effort to improve elections over the last 20 years. The Florida voting debacle in the 2000 presidential election (Mebane, 2004; Wand et al, 2001) led the United States to allocate billions of dollars (Congressional Research Service, 2021) to help fix problems with voting systems in the United States. One of the significant outcomes of this funding was the replacement of punch card and lever machines (as well as many paper ballot systems) with electronic voting systems, known as DREs (direct recording electronic voting devices). Not long after the transition to DREs, voting security experts began to question whether electronic voting machines that had no paper trail could be trusted, since there was no way to accurately verify that the machine had recorded and counted the votes as they had been cast. From these concerns, a new voting technology, known as the Ballot Marking Device (BMD) was introduced. The main difference between a DRE and a BMD is that BMDs do not record the vote electronically, and the paper ballot produced is the only official ballot. The BMD retains many of the desirable characteristics of a DRE, such as accommodating voters with physical disabilities, and preventing certain kinds of voting errors from occurring (e.g., stray ballot marks, voting for more than one candidate) while having the additional benefit of having an auditable paper trail of election results.

The use of some form of paper ballot in the election process is making a resurgence, and the percentage of jurisdictions having some form of paper record has increased from 71% in 2016 to 88% in 2020 (Verified Voting, 2020). Although only 19.6% of registered voters currently live in jurisdictions that vote exclusively with BMD's, there is a strong upward trend in their use, with an 11-fold increase since 2018 (Verified Voting, 2020).

As BMD use has increased, some voting experts maintain that these types of voting systems pose an unacceptable security risk to elections (Appel et al, 2020, Stark & Xie, 2019). One of the major concerns raised by Appel and his colleagues is that even though a BMD produces a paper record, the BMD itself may be compromised by malicious attackers. In the scenario envisioned by these researchers, the ballot marking device would show the voter the choices they had made on the screen, but would then alter those choices on the printed ballot. A voter who did not carefully check their printed ballot might then be unaware that their choices had been altered and the cast paper ballot would not reflect the will of that voter. These concerns have found their way into the US court system, with one prominent case in Georgia that had asked the courts to require a switch to hand marked ballots still being unresolved (Huseman, 2022).

Much of this concern about the inability of voters to detect changes to their printed paper ballots comes from previous research (Acemyan, et al., 2013; Bernhard, et al., 2020). Research into review screens on electronic ballots found that those voters also had difficulty detecting changes (Campbell & Byrne, 2009; Everett, 2007). In each of these studies, the measure of detection was calculated for *all* voters who participated in the study, regardless of whether or not they actually inspected the ballot or review screen for anomalies. The resulting data indicated that voters were generally not successful at detecting these kinds of changes, and researchers focused on trying to determine how to best overcome the suggested cognitive and perceptual limitations of voters in order to mitigate these low detection rates.

However, a recent paper by Kortum, Byrne, and Whitmore (2020) examined the data in a more refined way, measuring the detection rate for only those voters who actually inspected their ballots. Analyzing the data in this way showed that 76% of voters could detect anomalies on their printed ballots. Unfortunately, this original study was not specifically designed to examine this particular condition, so the statistical power of the analysis was low, even though the results were highly suggestive that voters actually could detect these kinds of changes. The current paper rectifies this problem by describing the results of a study where voters were specifically instructed to examine their ballots for accuracy after they had voted.

**Method**

**Participants**

This study utilized a total of 64 participants who were eligible to vote in the United States. At the time of the experiment, all participating voters resided in Houston, Texas, and were recruited using advertisements and referrals. In order to qualify for the study, participants were required to be at least 18 years of age, read and speak English, have normal or corrected-to-normal vision, and have normal motor capabilities. Of the 64 total participants, 31 were male and 33 were female. Participants ranged from 18 to 69 years of age, with a median age of 21 years. The study involved a diverse pool of participants, with 28.1% identifying as African American, 26.5% as Caucasian, 21.9% as Hispanic/Latino/Mexican American/Chicano, 18.8% as Asian American, 3.1% as Multicultural, and 1.6% as Other. Participant education levels reflected similar diversity. 26.7% of participants had received either a high school diploma or G.E.D., 48.4% reported some college or an associate degree, 18.7% had a bachelor's degree, and 6.2% had a graduate degree. All participants were compensated with $25 Amazon gift cards that were emailed to them following the completion of their task. The vast majority of the participants (93.8%) had voted at least once on an electronic voting machine, 37.5% had utilized paper ballots marked by hand, 3.1% had used lever machines, and 6.25% had voted with punch cards. Overall, 87.5% participants had voted in at least one national election and 59.4% had participated in local elections.

**Experimental Design**

This experiment employed a 2 (printed ballot design) x 2 (ballot length) x 2 (number of changes) x 2 (location of changes) between-subjects design.

*Printed ballot design:* The layout of the electronic ballot was unchanging across conditions; however, the printed paper ballot layout was split between two types. The first ballot was based on the Election Systems & Software (ES&S) design (Figure 1A) (Election Systems and Software, 2020). The second printed ballot design was modeled after the VSAP system (Figure 1B) (LA County Registrar-Recorder /County Clerk, 2021). These two ballot layouts were chosen with the purpose of determining whether there would be a significant difference in readability between the traditional design (ESS) and one crafted with user-centered design principles (VSAP).

*Ballot length:* Two ballot lengths were used in the study; a ballot with 40 individual contests and a ballot with 5 individual contests. These ballots are based on average ballots lengths across all 50 states (Ballotpedia, 2018), plus and minus one standard deviation, then rounded for easier experimental administration.

*Number of changes:* Two types of ballot selection changes were explored: a single targeted change and multiple changes (two changes for the 5-contest ballot and ten changes for the 40-contest ballot). The one change condition was set to resemble a targeted attack on a particular contest, for instance, a high profile office such as the presidency. The goal of the multiple change condition was to imitate broader attacks or down-ballot attempts to impact an election.

*Location of changes:* Two locations of changes were chosen for this study: the beginning part of the ballot (top 25%) or the middle part of the ballot (middle 50%). In the one-change top-of-ballot condition, the change occurred with the presidential contest, while in the middle-of-ballot condition, the change occurred in a random contest in the middle 50% of the ballot. For the multiple-changes, top-of-ballot condition, the changes included the presidential contest along with other random contests in the top 25% of the ballot. In the multiple-changes, middle-of-ballot condition, all changes occurred with random contests in the middle.

**Materials**

To better replicate experiences with an electronic voting machine, an 18-inch computer touchscreen tablet was utilized. The screen was installed into a frame box that angled and positioned the touchscreen to better resemble touch-screen voting machines. This setup was surrounded by voting-themed privacy screens to make the system appear as it might in a typical polling center. Qualtrics was employed to craft the electronic ballot and overall ballot interface design. This interface was designed to be similar to the VSAP system (Figure 2).

This study utilized the same names on the ballots that Kortum, Byrne, and Whitmore (2020) used in their study. While previous studies (Acemyan, Kortum & Payne, 2013; Campbell & Byrne, 2009; Everett, 2007) have used computer-generated names on the ballots, in this study we used widely recognizable names that are apolitical, such as Indiana Jones, Thomas Edison, and Amilia Earhart, so that participants might more easily remember these names when looking for ballot anomalies, as they would in an actual voting situation. This may still retain some of the important beneficial characteristics of using fake names, such as allowing participants to keep their political preferences private and not putting them in a situation where they might have to vote in a race where they hold an opinion (Quesenbery and Chisnell,2009). Participants did not have to pick candidates from these highly recognizable names, but rather were given prepared lists (slates) of candidates they were to select, in order to have greater control over the experiment, (e.g., Campbell, Tossell, Byrne, & Kortum, 2014; Greene, Byrne, & Everett, 2006; Redish, Chisnell, Laskowski, & Lowry, 2010). Participants were allowed to use their slates throughout the study.

A wizard-of-oz protocol was utilized for the ballot printing process. A wizard-of-oz protocol (Dahlbäck, 1993) means that participants are led to believe that they are using a fully realized device, while in fact the experimenters are manipulating certain aspects of the machine unbeknownst to participants. In the experiment, voters believed they were printing their ballots when they clicked the "PRINT" button, when in truth, the experimenters were controlling the printing process in order to control the number of ballot changes that were presented to the voters. Because the ballots were pre-generated, how closely participants followed their slates was not measured.

**Procedure**

Participants were first instructed to fill out the date, their names, and their emails to ensure they received their compensation following the completion of their tasks. Following this, participants reviewed and completed an Institutional Review Board (IRB)-approved consent form. Next, participants were directed toward the voting machine and given a 5 or 40 contest slate (depending on the condition) that they were instructed to use as a guide with which to vote, and given verbal instructions to check their printed ballot when they were done voting. Each participant was randomly assigned a condition that decided the type of ballot they would receive. Before participants voted, they were told "After you vote, the voting machine will produce a paper ballot. Your job is to identify anywhere where the paper ballot produced by the voting machine differs from how you voted." Once they had finished making their selections, they would then click the "PRINT" button and the experimenter would simultaneously print the manipulated ballot. Participants were then told to determine if there were any discrepancies between their intended candidates and the candidates on their ballots with the instruction "The next thing we'll have you do is check the printed ballot for errors. On the ballot, circle/highlight any errors you find. There may or may not be any errors on your ballot." Following this, participants were then tasked with completing a survey that covered information concerning their opinion on the overall usability of the voting system, their voting history, and demographics.

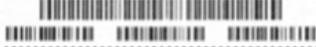
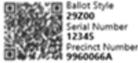
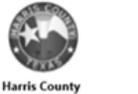
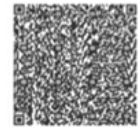

**A**  **B**

Figure 1: (A) ES&S-style printed ballot form. (B) VSAP-style printed ballot form. Both ballots show the long ballot with checking instructions.

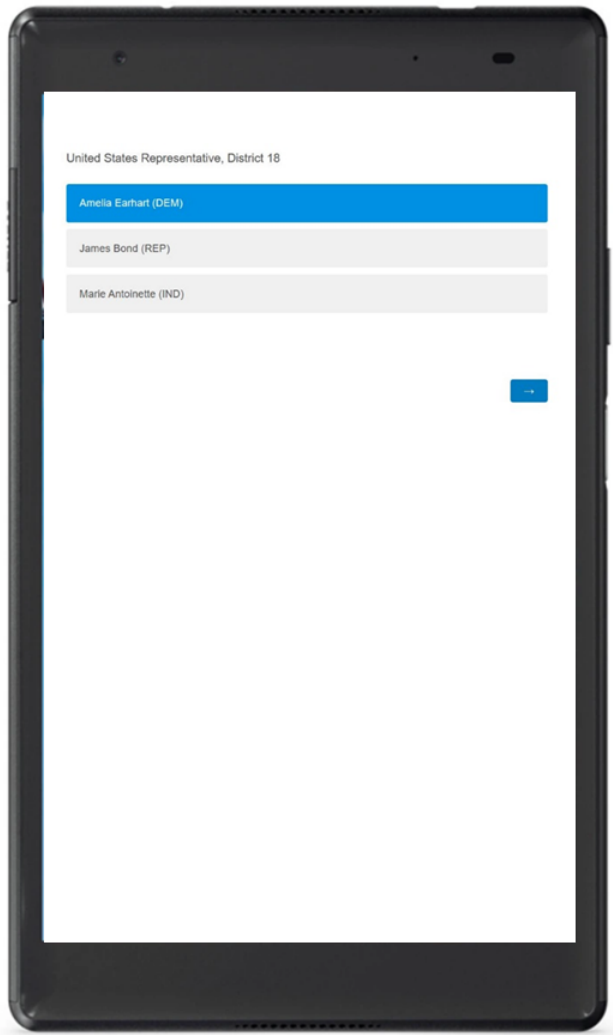

Figure 2: A representative vote selection screen on the ballot marking device.

**Results**

Overall accuracy was exceptionally high. Subjects saw 1440 total contests, and of those 1440, there were a total of 4 errors, so total accuracy was 99.8%. 3 of those errors were misses, that is, where a candidate had been changed and the subject did not flag it, and one error was a false alarm, where a candidate was not changed and the subject flagged it as having been changed.

As a result of the near-zero variability in performance, none of the independent variables had any effect on detection performance. That is, subjects were able to perform with near-perfect accuracy regardless of ballot length, ballot type, number of flipped races, and location of flipped races. Detection performance was *extremely* robust.

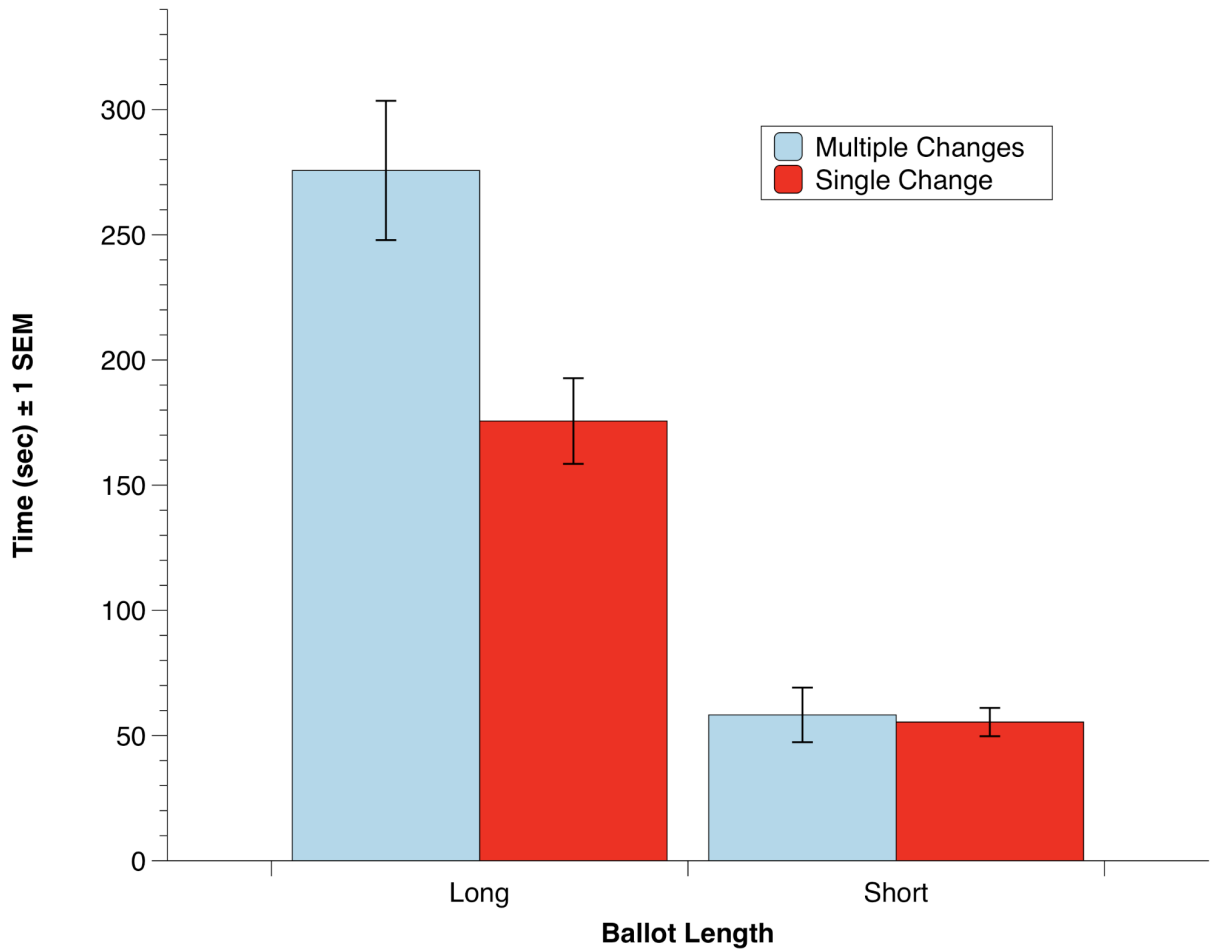

Figure 3. Time to review the ballot as a function of ballot length (short = 5 contests, long = 40 contests) and number of flipped votes

Time taken to perform the task was also recorded, and here there was variability. Figure 3 presents mean time to review the ballot as a function of ballot length and the number of changes made. Clearly—and not surprisingly—longer ballots took longer to review. The difference between long and short ballots was considerable; main effect of ballot length $F(1, 48) = 96.12$, $MSE = 4745$, $p < .001$, Cohen's $f = 1.42$. Note that the effect of ballot length was quite large not only in the statistical sense, but in practical terms as well; the overall average time for a 5-race ballot was under a minute (57 sec) and for 40-race ballots it was almost four minutes (226 sec). Again, it is not surprising that the long ballots took longer given that they contained eight times as many contests. The nearly-perfect accuracy and increased time for longer ballots clearly indicates that people are able to invest time in order to maintain high levels of performance. People can do this task highly accurately if they take the time to do so.

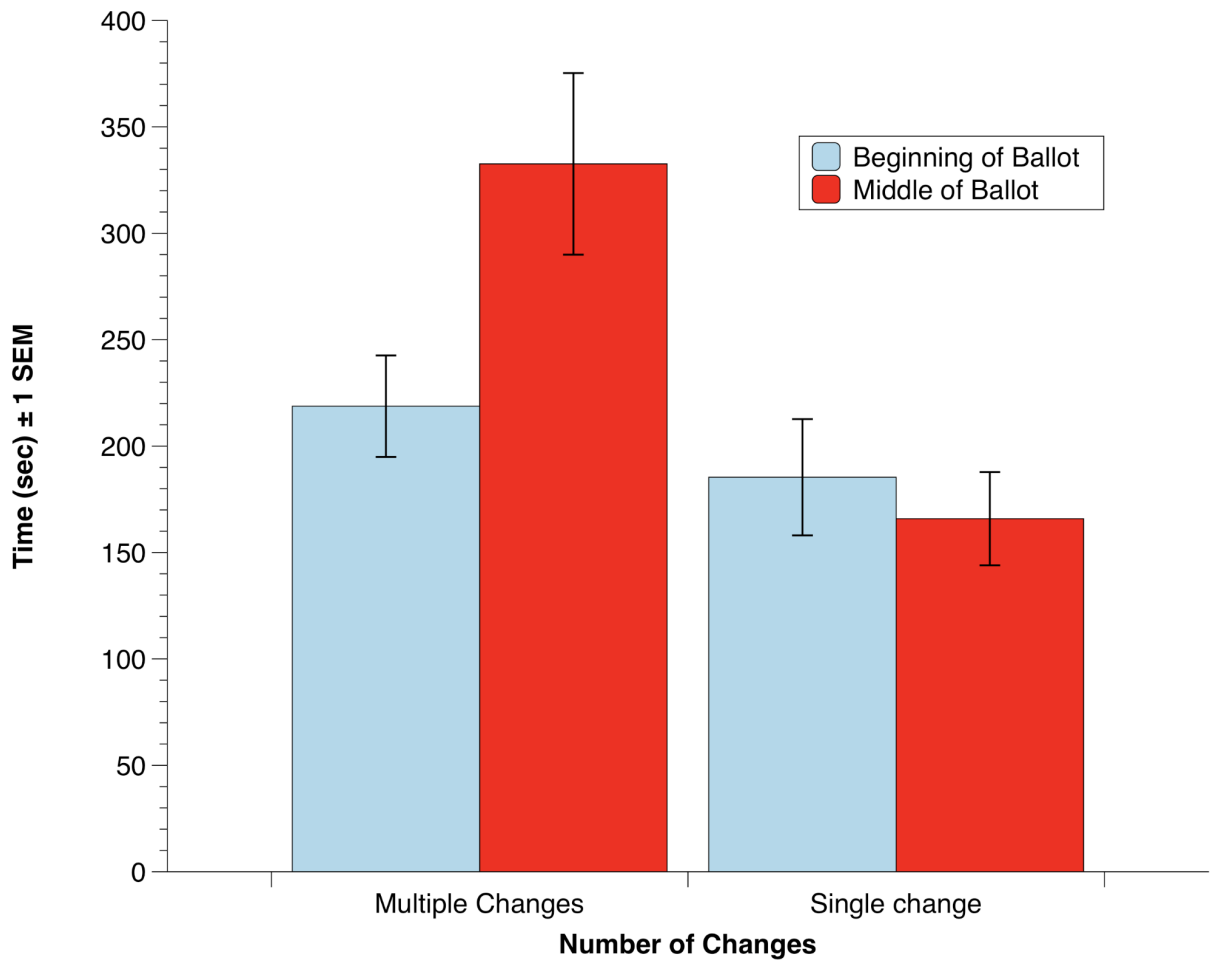

Figure 4. Ballot review time for long ballots as a function of the number of changes and location of the changes.

There was also a three-way interaction between ballot length, number of changes, and location of changes, $F(1, 48) = 4.66$, $MSE = 4745$, $p = .036$, Cohen's $f = 0.31$. While often three-way interactions can be difficult to interpret, this one is actually straightforward. For short ballots, time taken was insensitive to the other independent variables. However, for long ballots, there was a two-way interaction between number of changes and change location, as shown in Figure 4; $F(1, 24) = 4.32$, $MSE = 8237$, $p = .048$, Cohen's $f = 0.42$. This interaction is essentially driven by the fact that the time for multiple changes in the middle of the ballot was longer than the other three conditions.

It is not entirely clear what this result means in practical terms. The fact that changes in the middle of the ballot take longer to detect suggests that when voters review, they start at the top of the ballot, but that would imply that finding a single error mid-ballot should take longer than finding a single error at the top of the ballot, which was not what was found. Further research will be necessary to clarify the basis for this effect.

**Discussion**

Critics of BMDs have claimed that BMDs are in-principle insecure because voters are incapable of accurately detecting modifications of their choices. However, there is a critical difference between *cannot* and *do not*; prior research did not carefully differentiate between these. As more recent research has shown, the low detection rate shown in early research was actually due to the way in which the data were being analyzed. It is clear from the current results and other recent research that there are no intrinsic perceptual or cognitive limitations that cause voters to perform poorly at this task. Rather, provided the right direction and support, voters are almost perfect detectors of changes in ballots. This has important implications for the widespread deployment of ballot marking devices, and suggests that much of the human security necessary to catch errors is already in place, if it can be effectively utilized.

The other factor that makes this result so remarkable is that there was nothing particularly remarkable about the pool of voters. These were not people who were recruited on the basis of any kind of relevant work or life experience (such as proofreading or auditing), and they were given no special training. Nor was there any evidence that any demographic variable (e.g., level of education) had any impact on detection performance. While it is impossible to guarantee this high a level of performance from all voters, the fact that this otherwise unremarkable sample yielded such a strong result suggests that this phenomenon is highly robust.

It is important to note that voters in this experiment were given a printed list of who they were supposed to vote for. This almost certainly facilitates detection performance because it circumvents potential memory failures on the part of voters. That is, the task here was a relatively straightforward visual comparison task. Nonetheless, it is not a given that even with such an aid that voters would perform so well. While many voters do not bring a "cheat sheet" into the polling place with them, there is no barrier to this in most jurisdictions. Our results demonstrate what is possible.

The question moves away from voters' capabilities to "How do we best motivate and incentivize voters to check their ballots before depositing them in the ballot box?" This is a more straightforward administrative question about providing opportunities for the voter and creating processes that encourage this important behavior. The challenge is now to research and devise these effective administrative processes that support the voters in this critical task. What might these administrative processes look like?

*Signage inside the polling place*. Perhaps the most obvious place to start is by simply alerting the voter in the polling station that the paper ballot is the ballot of record and that they need to carefully check it before depositing it into the ballot box. These admonitions could take place as voters enter the polls, they could be printed on the ballots themselves and there could be reminders on the BMD screens. Previous research on these kinds of instructions/warnings shows that they are easy to ignore (Wright, 1981), and counterintuitively, the more frequent warnings are, the less likely it is that users will heed those warnings (Ramsay, 1989). Results from Bernhard (2020) showed limited effectiveness in the use of these warnings, but a systematic exploration of warning language and placement might find a more effective combination.

*Public service/social media campaigns prior to the election.* A similar strategy is the use of public service announcements and social media campaigns prior to and on the day of the election, telling voters that they are the last line of security defense in the election, and that checking their ballot is the best way to help ensure a secure election. While more difficult to examine experimentally, there have been studies that have explored the use of public service announcements and social media campaigns in order to alter voter behavior. (Burgess et al., 2000; De Rooij et al., 2017). Perhaps this, combined with the signage inside the polling station, would prove to be effective.

*Ballot checking stations.* A more significant change, but one that might produce greater effects, would be the implementation of ballot checking stations. Here the physical layout of the polling station would be changed so that after voters had received their printed ballot from the BMD they would then proceed to a specific ballot checking station. The purpose of this station would be twofold. First, it would allow the voter the opportunity to examine their ballot carefully, without the pressure of having other voters waiting for the voting machine to be freed up. During peak periods, voters may feel intense pressure to vote and leave the ballot marking device quickly. By providing a separate station for checking the ballot, voters might be willing to spend additional time. Second, it would formalize the process of checking the ballot. Now, the mental model of the voter would be "this is the station where I vote" (the BMD), "this is the station where I check my ballot for errors" (the ballot checking station), and "this is the station where I cast my checked ballot" (the physical ballot box). This could give the voter an incentive to perform the checking task, since it is now simply part of the three stage process. Of course, understanding how this might impact polling station space requirements, voter throughput, and wait times in polling stations would be paramount in this kind of polling place alteration.

*Gamification of the process.* If these standard administrative steps demonstrate themselves to be ineffective in shaping voter behavior, more radical approaches might need to be considered and researched. For example, gamification (Deterding et al., 2011) elements could be built into the system that would promote voters spending more time looking at their ballots, and potentially being rewarded for that checking behavior. The exact details of how this might work would need to be seriously examined, but it is a strategy that has proven effective in other domains, such as health and consumer business operations (Pereira et al., 2014; Robsin et al., 2016) and is now being considered in more public service domains as well (Asquer, 2013).

Each of these scenarios, and the impact they have on the voter and the effective operation of a polling station, should be investigated in future research. Since these changes would largely be administrative, expensive modifications or purchases of equipment/software would be unnecessary, suggesting that enhanced security could be obtained with minimal additional expense.

Finally, we remain in agreement with critics of BMDs that it should not necessarily be the voters who bear full responsibility for voting system security. Voters should be encouraged to bring cheat sheets and thoroughly check their ballots to protect themselves from their own potential mistakes, regardless of security issues. While having thorough voters is desirable, it should not be the only defense. Thus, we encourage election officials to not only explore ways to encourage voters to check their printed ballots, but to conduct live audits of BMDs. Multiple layers of defense are critical to ensuring election integrity.

**Conclusion**

The data presented in this paper show that, with proper direction, voters can be near-perfect detectors of ballot changes on printed paper ballots after voting with a BMD. This finding has significant implications for the voting community as BMD use continues to grow. The key now is to identify administrative and behavioral methods that will prompt and encourage voters to check their BMD-generated ballots before they drop them in the final ballot box.

**Acknowledgments**

This work was supported in part by NSF grant numbers SMA-1853936 and SMA-1559393 and in part by a grant from the William and Flora Hewlett Foundation through the MIT Election Lab. The views and conclusions contained herein are those of the authors and should not be interpreted as representing the official policies or endorsements, either expressed or implied, of